\documentclass[symmetry,article,accept,moreauthors,pdftex]{Definitions/mdpi} 

\firstpage{1} 
\pubvolume{xx}
\issuenum{1}
\articlenumber{1}
\pubyear{2020}
\copyrightyear{2020}
\history{Received: 27 February 2020; Accepted: 30 March 2020; Published: date}
\updates{yes} 

\usepackage{comment}

\Title{High-Resolution Radio Observations of Five Optically Selected Type 2 Quasars}

\Author{M\'at\'e Krezinger $^{1,2}$\orcidA{}, S\'andor Frey $^{2,3}$\orcidB{}, Zsolt Paragi $^{4}$\orcidC{} and Roger Deane $^{5,6,}$*\orcidD{}}

\AuthorNames{M\'at\'e Krezinger, Zsolt Paragi, S\'andor Frey and Roger Deane}

\address{%
$^{1}$ \quad \textls[-15]{Department of Astronomy, E\"otv\"os Lor\'and University, P\'azm\'any P\'eter s\'et\'any 1/A, H-1117 Budapest,~Hungary} \\
$^{2}$ \quad Konkoly Observatory, Research Centre for Astronomy and Earth Sciences, Konkoly Thege Mikl\'os \'ut 15-17, H-1121 Budapest, Hungary; frey.sandor@csfk.mta.hu\\
$^{3}$ \quad Institute of Physics, ELTE E\"otv\"os Lor\'and University, P\'azm\'any P\'eter s\'et\'any 1/A, H-1117 Budapest, Hungary\\
$^{4}$ \quad Joint Institute for VLBI ERIC, Oude Hoogeveensedijk 4, 7991 PD, Dwingeloo, The Netherlands; zparagi@jive.eu\\
$^{5}$ \quad Department of Physics, University of Pretoria, Hatfield, Pretoria 0028, South Africa; roger.deane@up.ac.za \\
$^{6}$ \quad Centre for Radio Astronomy Techniques and Technologies, Department of Physics and Electronics, Rhodes University, Mkhanda 6140, South Africa}

\corres{Correspondence: krezinger.mate@csfk.mta.hu}

\abstract {Many low-luminosity active galactic nuclei (AGNs) contain a compact radio core which can be observed with high angular resolution using very long baseline interferometry (VLBI). Combining~arcsec-scale structural information with milliarcsec-resolution VLBI imaging is a useful way to characterise the objects and to find compact cores on parsec scales. VLBI imaging could also be employed to look for dual AGNs when the sources show kpc-scale double symmetric structure with flat or inverted radio spectra. We observed five such sources at redshifts $0.36<z<0.58$ taken from an optically selected sample of Type 2 quasars with the European VLBI Network (EVN) at 1.7 and 5 GHz. Out of the five sources, only one (SDSS~J1026--0042) shows a confidently detected compact VLBI core at both frequencies. 
The other four sources are marginally detected at 1.7~GHz only, indicating resolved-out radio structure and steep spectra. Using first-epoch data from the ongoing Karl G. Jansky Very Large Array Sky Survey, we confirm that indeed all four of these  sources have steep radio spectra on arcsec scale, contrary to the inverted spectra reported earlier in the literature. However, the VLBI-detected source, SDSS~J1026$-$0042, has a flat integrated spectrum. Radio AGNs that show kpc-scale symmetric structures with truly flat or inverted spectra could still be promising candidates of dual AGNs, to be targeted with VLBI observations in the future.}

\keyword{active galactic nuclei; quasars; radio continuum; radio spectrum; interferometry; imaging}

\begin{document}

\section{Introduction}
\label{s:intorduction}

Despite decades of detailed investigation, the~true nature of the low-luminosity active galactic nuclei (LLAGNs, e.g.,~Seyferts and LINERs) is still under debate. More than 40\% of galaxies in the local Universe contain LLAGN~\cite{ho1997}. Since they are much fainter (with bolometric luminosity $L_{\rm{bol}} < 10^{38}$~W and radio power $P_{\rm{1.4GHz}} < 10^{35}$~W\,Hz$^{-1}$) than the radio-loud AGNs and typically lack prominent, extended jets, it is more difficult to observe them. As~with all AGNs, these sources are also powered by accretion onto a supermassive black
hole, but~with a lower accretion rate. Thus they appear as ``downscaled'' AGNs (e.g., \cite{falcke2000}). These sources show mostly low ionization state optical emission lines in their spectra~\cite{ho1997}, and~have low accretion rates ($< 10^{-2}-10^{-3}$ times the Eddington rate~\cite{nagar2002}). LLAGNs~also have radio emission that seems originating from compact jets on parsec (pc) scales~\cite{falckebiermann1999, nagar2002, nagar2005}.  

Using the technique of very long baseline interferometry (VLBI), we can directly observe compact pc or even sub-pc scale radio emission. The~baselines of the globally distributed radio telescope arrays, such as the European VLBI Network (EVN), are up to a few thousand kilometers long, providing milliarcsec (mas) or sub-mas resolution, depending on the observing frequency. Because~of the high resolution, a~VLBI detection of a radio source at GHz frequencies typically implies high brightness temperatures ($T_{\rm b} > 10^{6-7}$ K) that can be produced by the non-thermal (synchrotron) emission of AGNs~\cite{kewley2000, herrera2016}. 
A radio power exceeding $10^{21}$~W\,Hz$^{-1}$ in a source unresolved by the Karl G. Jansky Very Large Array (VLA) on arcsec scales is already indicative of accretion-powered nuclear activity~\cite{kewley2000} or~maybe circumnuclear starburst. Detection of a compact core with VLBI in a high-luminosity source is a very strong AGN~indicator.

In LLAGNs, compact radio emission with flat (i.e.,  with~spectral index $-0.5 \le \alpha \le 0$, where~the flux density is $S \propto \nu^{\alpha}$, and~$\nu$ is the frequency) or inverted ($\alpha > 0$) radio continuum spectrum can indicate radio emission that is fueled by a synchrotron self-absorbed jet base coupled to an underluminous accretion disk (e.g., \cite{falckebiermann1999}) or an accretion inflow onto the black hole~\cite{narayan1998}. When observed at high frequencies, magnetized plasma in the accretion disc corona can also produce compact sub-pc scale optically thick flat-spectrum radio emission (\cite{panessa2019} and references therein). On~the other hand, external free--free absorption of non-thermal radio emission can cause flat spectrum by flattening the steep-spectrum ($\alpha < -0.5$) external emission~\cite{ulvestadho2001,lalandho}, although~it would have brightness temperature $T_{\rm b} \sim 10^{4}$ K~\cite{condon1991}. There is a phenomenon analogous to coronal mass ejections in active stars that could also occur in AGNs, in~the form of outflowing blobs of highly magnetized plasma~\cite{laor2019}. However,  similar to  jet activity, it produces extended optically thin emission with steep radio spectrum. At~highly subluminal speeds, the~magnetized corona could be considered as the base for jet launching (\cite{panessa2019} and~references therein).

The combination of radio interferometric measurements with the VLA and VLBI is an effective tool to identify compact cores in LLAGNs (see~\cite{nagar2002, nagar2005} and references therein). At~GHz frequencies, obscuration~effects are minimal compared to the infrared or ultra-violet regimes, and~sensitive high-resolution radio maps of a source can be produced with typically less than 1--2~h of observation. The~arcsec-scale {{multi-frequency}} VLA imaging can provide essential spectral and structural information~\cite{fu2015,lacy2020}, which can help selecting targets for follow-up VLBI observations of the mas-scale radio structure. With~this much finer angular resolution, the~extended radio emission from the host galaxy is resolved out in general, thus only the compact emission from the AGN remains to be detected. {AGN cores typically have flat radio spectra, as~opposed to extended jet or hot spot features. Therefore,   arcsec-scale radio imaging is a useful tool to identify candidate multiple AGN systems. Sources with multiple flat-spectrum features can then be targeted with VLBI observations that provide mas-scale resolution. A~compact, high brightness temperature ($T_{\rm b} > 10^{6}$~K) component that can only be revealed with VLBI is a clear indication of AGN core emission (\cite{an2018,derosa2020}  and~references therein).}

This method could offer an excellent means for finding multiple AGNs with kpc separation. Dual~AGNs are expected at stages of galaxy evolution driven by mergers throughout the history of the Universe. While their observational identification is challenging, their study is important in many areas of current cosmology and multi-messenger astrophysics (see~\cite{derosa2020,an2018}  and references therein). VLBI observations of a large sample of arcsec-scale double radio AGNs with flat/inverted overall spectra would provide an opportunity to estimate the prevalence of dual radio-emitting AGNs in an unbiased way. Synchrotron self-absorbed jets, and~ emission from a magnetically-heated corona in an accretion disc can show flat/inverted spectrum~\cite{laorbehar2008,bondiperez2010}. In~either case, the~compact double structure would indicate dual AGN activity since the detection of mas-scale emission at GHz frequencies is directly related to AGN activity in sources above redshift $z\approx0.1$ \cite{derosa2020}. As~compact cores are relatively common in LLAGNs, they are good candidates for finding dual AGNs. This would provide an efficient selection method that can be applied for future sub-arcsec resolution interferometer surveys,  such as  the ones using the SKA~\cite{deane2015}. {Securely confirming dual AGN candidates is not a trivial task that can solely rely on radio measurements. It often benefits from observations at multiple wavebands, from~optical to X-rays, using both spatially resolved spectroscopic and imaging techniques~\cite{ellison2017,derosa2020}.}

In this paper, we report on the observations of five LLAGNs with the EVN at two frequencies, 1.7 and 5 GHz. Section~\ref{s:sample} presents the sample and its selection. Section~\ref{s:observations} gives details about the radio interferometric observations. The~data analysis and the result are described in Section~\ref{s:datared}, and~our findings are discussed in Section~\ref{s:discussion}. We conclude the paper in Section~\ref{s:conclusions}. Throughout this study, we adopt a standard flat $\Lambda$ Cold Dark Matter cosmology with $\Omega_{\rm{m}} = 0.3$, $\Omega_{\Lambda} = 0.7$, and~$H_{\rm{0}} = 70$~km\,s$^{-1}$\,Mpc$^{-1}$ for calculating luminosities and projected linear sizes~\cite{wright2006}.
 
\section{Sample Selection and Observing~Goals}
\label{s:sample}

\textls[-25]{We selected five sources for mas-resolution follow-up VLBI observations from the sample of~\cite{lalandho}. Our~target sources are listed in Table~\ref{tbl:sources}. These were previously observed in the radio at 8.4~GHz with the VLA in its B configuration~\cite{lalandho}, providing typical angular resolution of about 1~arcsec or somewhat below. Originally, the~sources were taken from the optically selected sample of \citet{zakamska2003}, which~was the first comprehensive survey for studying the optical properties of Type 2 quasars (i.e.,~quasars where the accretion disk is seen nearly ``edge-on'', and~the disk and the broad-line region are obscured by the surrounding torus of dense gas and dust) in detail using SDSS data. Each of the five sources we selected (Table~\ref{tbl:sources}) showed symmetric double radio structure on arcsec angular scales in the 8.4-GHz VLA images, and~their overall radio spectra were reported to be inverted~\cite{lalandho}. Two of our targets, J0741$+$3020 and J0956$+$5735, were included in the sample of \citet{bellocchi2019}, who looked for ionized gas outflows in obscured quasars. According to their optical spectroscopic observations, outflows are detected in both objects. J0741+3020 is associated with a giant ($>100$~kpc) ionized nebula, while in the case of J0956+5735, the~compact [OIII] profile suggests a spatially unresolved outflow~\cite{bellocchi2019}. The~host galaxies of three of these AGNs are morphologically classified from SDSS or {\em Hubble Space Telescope} images as bulge-dominated elliptical/S0 galaxies (Table~\ref{tbl:sources}), with~J0741$+$3020 as an ULIRG elliptical in a multiple system~\cite{bellocchi2019,villarmartin2012}. J1447$+$0211 is difficult to classify as it appears highly disturbed due to a recent galaxy interaction~\cite{urbanomyorgas2019}. It shows optical features  such as tidal tails, bridge, and~even double nuclei with 1.5~kpc separation. For~the host galaxy of J1026$-$0042, no classification is found in the~literature.}

\begin{table}[H]
\caption{\textls[-15]{Name, J2000 optical equatorial coordinates, redshift, VLA 8.4-GHz radio flux density, 1.4--8.4 GHz spectral index, and~optical magnitude of the five sources selected from \citet{lalandho}. The~coordinates are taken from the SDSS Data Release 15~\cite{sdssdr15} and, for~SDSS~J1026$-$0042, the~{\em Gaia} Data Release 2~\cite{gaia2016,gaia2018}. The~SDSS $r$-band magnitudes are adopted from \citet{zakamska2003}. The~morphological types of the visually classified host galaxies are obtained form the works in~\cite{bellocchi2019,villarmartin2012,urbanomyorgas2019}. HD means highly disturbed from galaxy~interaction}.}  
\label{tbl:sources}
\centering 
\begin{tabular}{cccccccccccc}
\toprule
\multirow{2}{*}{\textbf{ID}} & \multicolumn{3}{c}{\textbf{Right} \textbf{Ascension}} & \multicolumn{3}{c}{\textbf{Declination}} & \multirow{2}{*}{\boldmath{$z$}} & \boldmath{$S_{\rm{8.4\,GHz}}$} & \multirow{2}{*}{\boldmath{$\alpha^{\rm{8.4}}_{\rm{1.4}}$}} & \boldmath{$r$} & \textbf{Host}\\
   &\textbf{ h} & \textbf{min} & \textbf{s} & \boldmath{$^\circ$} & \boldmath{$^\prime$} & \boldmath{$^{\prime\prime}$}&  & \textbf{mJy} & & \textbf{mag} & \textbf{Galaxy} \\
\midrule
 SDSS~J0741$+$3020 & 07 & 41 & 30.513 & $+$30 & 20 & 05.186 & 0.476 & 2.97 & $+$0.56 & 21.06 & ULIRG \\
 SDSS~J0956$+$5735 & 09 & 56 & 29.058 & $+$57 & 35 & 08.775 & 0.361 & 2.88 & $+$0.72 & 20.18 & E/S0 \\
 SDSS~J1014$+$0244 & 10 & 14 & 03.524 & $+$02 & 44 & 16.521 & 0.573 & 4.53 & $+$0.45 & 21.51 & E/S0\\
 SDSS~J1026$-$0042 & 10 & 26 & 40.437 & $-$00 & 42 & 06.484 & 0.364 & 3.68 & $+$0.38 & 19.73  & ...\\
 SDSS~J1447$+$0211 & 14 & 47 & 11.292 & $+$02 & 11 & 36.234 & 0.386 & 2.76 & $+$0.85 & 20.35 & HD\\
\bottomrule
\end{tabular}
\end{table}

Our sources selected for VLBI observations (Table~\ref{tbl:sources}) fall in the radio-intermediate regime, between~the radio-quiet and radio-loud AGNs. Interestingly, most of the sources in the 8.4-GHz VLA sample~\cite{lalandho} with flat or inverted spectra are radio-intermediate objects, and~some of them show extended
symmetric double structure on arcsec scale. A~flat/inverted spectrum may indicate free--free absorption by ionized gas in the narrow-line region. This finding was rather unusual, which prompted us to probe the structure of our sub-sample of ``best established'' symmetric sources (Table~\ref{tbl:sources}) on mas scales with VLBI observations. Finding a mas-scale compact radio core with deep VLBI imaging could in principle pin down the location of the AGN, which has been found offset from the optical position in a number of cases (e.g., \cite{narayan1998}), while the extended radio
emission should be completely resolved out (undetected) on the long interferometer baselines. Although~an unlikely scenario for the given sample, one could even find instead two compact VLBI cores corresponding to dual AGN activity, given the symmetric flat/inverted components seen on VLA scales. In~this case, the~arcsec-scale symmetric double radio structure could be explained by simultaneous activity in both~AGNs.

\section{Observational~Data}
\label{s:observations}
\unskip

\subsection{EVN~observations}
\label{ss:evnobs}

We observed the five selected targets (Table~\ref{tbl:sources}) with the EVN at 1.7 and 5~GHz, to~attempt the detection of mas-scale compact radio components in these LLAGNs. The~observations were carried out on 5 June 2015 (5~GHz) and 15 June 2015 (1.7~GHz) under the project code EP093. The~total bandwidth of 128~MHz was divided into eight intermediate frequency channels (IFs) containing thirty-two 500-kHz wide spectral channels each, in~both left and right circular polarizations. The~following 13 radio observatories participated in the experiments: Jodrell Bank Mk2 25-m (Great Britain), Westerbork 25-m (The Netherlands), Effelsberg 100-m (Germany), Medicina 32-m, Noto 32-m (Italy), Onsala~25-m (Sweden), Sheshan 25-m (China), Toru\'n 32-m (Poland), Yebes 40-m (Spain), Svetloe 32-m, Zelenchuckskaya 32-m, Badary 32-m (Russia), and~Hartebeesthoek 26-m (South Africa). The~data were recorded at 1024~Mbit\,s$^{-1}$ rate and correlated at JIVE. We applied the technique of phase referencing~\cite{beasley1995} and subsequently pointed the radio telescopes to a nearby bright compact calibrator source and the given target, in~a repeated 5-min cycle. Within~each cycle, $\sim3.5$~min was spent on the weak target source. The~delay, rate, and~phase solutions derived for the phase-reference calibrators (J0741+3112, J0956+5753, J1015+0109, J1024$-$0052, and~J1440+0157) could later be interpolated and applied to the corresponding target source data, to~allow for longer coherent integration and thus an improved sensitivity for the weaker LLAGNs. The~calibrator--target separations were 0.87$^\circ$, 0.31$^\circ$, 1.65$^\circ$, 0.57$^\circ$, and~1.57$^\circ$, respectively. The~total on-source integration time for each target was $\sim$80~min. 

\subsection{VLA Sky Survey~Data} 
\label{ss:vlassdata}

\textls[-25]{We obtained the raw 8.4-GHz VLA data analyzed by \citet{lalandho} from the NRAO Archive ({\url{ http://archive.nrao.edu/}) (project code AV288) and calibrated them in a standard way using the NRAO AIPS software ({\url{ http://www.aips.nrao.edu/index.shtml}}), but~failed to reproduce the images of our five target sources.  
To obtain independent information on the overall radio spectrum of the sources, we~then turned to the publicly available preliminary radio images from the ongoing VLASS~\cite{lacy2020} at 2.7~GHz and compared the obtained flux densities to the 1.4-GHz values in the FIRST survey catalogue~\cite{becker1995}. The~VLASS supersedes previous VLA radio sky surveys at GHz frequencies  such as  the NVSS~\cite{condon1998} in terms of angular resolution and sensitivity. In~2019, the~first phase of the survey  was  completed with covering the entire sky north of $-40^{\circ}$ declination at the $2-4$~GHz frequency range in total intensity (Stokes I). The~angular resolution is 2.5$^{\prime\prime}$ and the imaging sensitivity is $\sigma \sim 120~\mu$Jy\,beam$^{-1}$ at the first survey epoch. Later, with~the planned three  epochs combined, the~sensitivity should reach $\sim$70~$\mu$Jy\,beam$^{-1}$. All our target sources fall into the VLASS coverage area. Quick Look images with $1^{\circ} \times 1^{\circ}$ field size are available in the VLASS archive ({\url{https://archive-new.nrao.edu/vlass/}}). Five~tiles (T18t11, T25t08, T11t16, T10t16, and~T11t23) were selected based on the celestial coordinates, corresponding to our targets in the order of their appearance in Table~\ref{tbl:sources}. The~Quick Look images are capable of revealing arcsec-scale morphological features of radio galaxies like radio cores and jets~\cite{hernandez2018}; however, they  may suffer from deconvolution issues due  to the extremely short integrations and resultant $(u,v)$ coverage.  }

\section{Data Analysis and~Results} 
\label{s:datared}
\unskip

\subsection{EVN~Data}

The raw correlated VLBI visibility data that are now in public domain and can be obtained from the EVN Data Archive ({\url{https://old.jive.nl/archive-info?experiment=EP093A_150605} and \url{https://old.jive.nl/archive-info?experiment=EP093B_150615}}) were loaded into AIPS for initial calibration. We followed standard procedures~\cite{diamond1995}. At~the beginning, the~data   affected by known problems with antenna pointing and radio frequency interference were flagged. The~visibility amplitudes were calibrated using antenna gains and system temperatures measured at the telescope sites. We used nominal system temperature values where measurements were unavailable. We corrected for the dispersive ionospheric delay using total electron content maps derived from global navigation satellite systems data, and~also applied parallactic angle corrections. Using a 1-min scan on the strong fringe-finder source 0528+134, we performed an initial correction of instrumental phases and delays. We then performed fringe-fitting for all the five phase-reference calibrator sources and the fringe-finders (0528+134, 3C\,138, and~4C\,39.25) scheduled in the experiments. These calibrated visibility data were exported to the Difmap package~\cite{difmap} for hybrid mapping. This involved iterations of the CLEAN algorithm and phase-only self calibration, and~finally phase and amplitude self-calibration. We used the brightness distribution models obtained for the calibrator and fringe-finder sources to derive overall antenna gain correction factors (up to 6\% in case of some antennas). These were applied in AIPS to the target source data as well, to~refine the initial amplitude calibration. As~the next step, we repeated fringe-fitting for the five phase-reference calibrators in AIPS. This time the CLEAN component models of their brightness distributions were used as an input, to~account for any residual phases due to their mas-scale structure. The~solutions were interpolated and applied to the data of the respective target~sources. 

\textls[-15]{The calibrated and phase-referenced visibility data of the targets (J0741$+$3020, J0956$+$5735, J1014$+$0244, J1026$-$0042, and~J1447$+$0211) were transferred from AIPS to Difmap for imaging. We~applied natural weighting with the visibility errors raised to the power $-1$ ({\tt uvweight 0,-1} command in Difmap) in order to minimize the image noise. At~both observing frequencies, 1.7 and 5~GHz, the~dirty images of all target sources but J1026$-$0042 showed peaks with signal-to-noise ratios below 7, as~seen in a rectangular field of view of 512~mas $\times$ 512~mas centered on the brightest pixel. Therefore,   we regard J0741$+$3020, J0956$+$5735, J1014$+$0244, and~J1447$+$0211 as non-detections with the EVN. For~J1026$-$0042, the~calibrated visibility data were fitted directly with a circular Gaussian brightness distribution model in Difmap~\cite{difmap}. The~images of J1026$-$0042 at both frequencies are displayed in Figure~\ref{fig:J1026}. }

\begin{figure}[H]
\centering
 \includegraphics[width=7 cm, angle=-90]{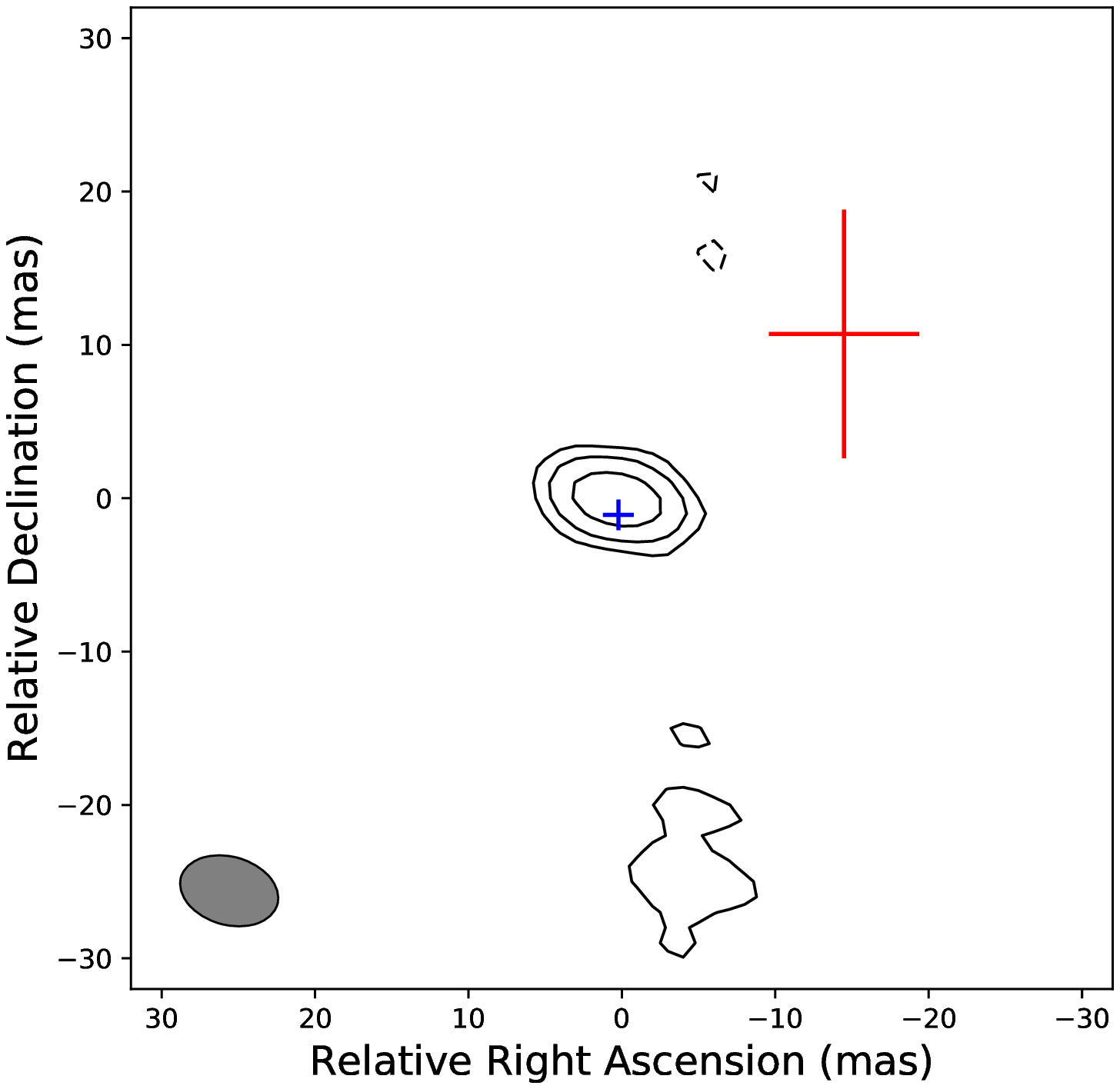}
 \includegraphics[width=7 cm, angle=-90]{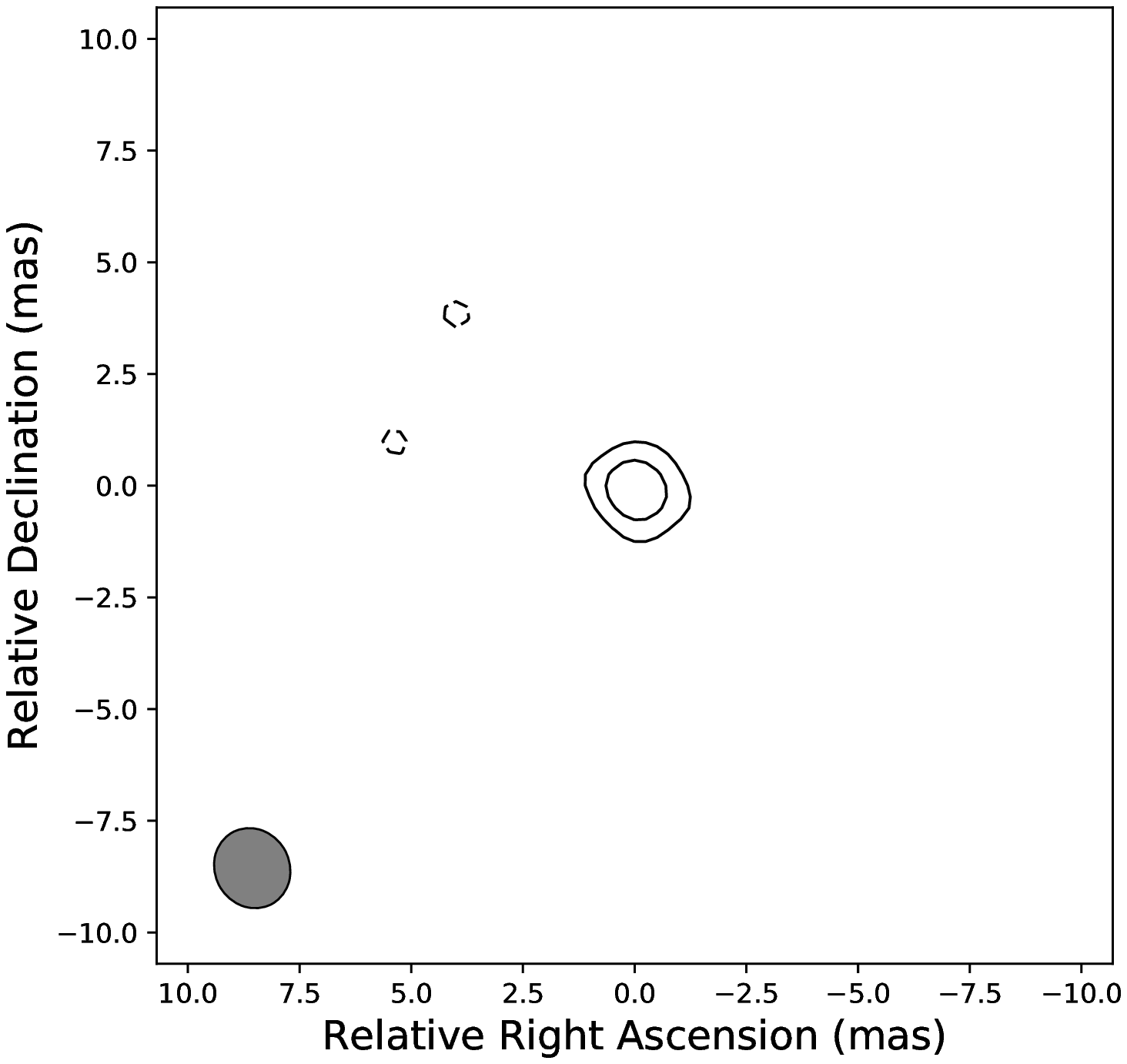} 
 \caption{VLBI images of SDSS~J1026$-$0042 at 1.7~GHz (left) and 5~GHz (right). The~red cross marks the {\em Gaia} DR2 optical position~\cite{gaia2016,gaia2018}, the~blue cross indicates the EVN 5~GHz position (see Section~\ref{ss:results-evn}). The~peak brightness values are 1.01~mJy\,beam$^{-1}$ at 1.7~GHz, and~0.20~mJy\,beam$^{-1}$ at 5~GHz. The~contour levels are drawn at 3~$\sigma~\times$ ($-$1, 1, 2, 4), where the 3~$\sigma$ image noise is $165~\mu$Jy\,beam$^{-1}$ and $65~\mu$Jy\,beam$^{-1}$ at 1.7 and 5~GHz, respectively. The~size of the elliptical Gaussian restoring beam (HPBW) is shown as an ellipse in the lower-left corner of the images. The~HPBW is 6.5~mas $\times$ 4.5~mas with a major axis position angle $ PA = 74.3^{\circ}$ (measured from north through east) at 1.7~GHz, and~1.83~mas $\times$ 1.67~mas with $ PA = 29.2^{\circ}$ at 5~GHz.}
 \label{fig:J1026}
\end{figure}

At 1.7~GHz, looking at the EVN dirty images of a larger ($4.096^{\prime\prime} \times 4.096^{\prime\prime}$) field of view, the~remaining four sources (J0741$+$3020, J0956$+$5735, J1014$+$0244, and~J1447$+$0211) have an SNR slightly above 7. Here, we also used natural weighting with the visibility errors raised to the power $-1$, except~for J0741$+$3020 and J1447$+$0211, where the power of $-2$ was applied in Difmap. Therefore,   these~LLAGNs are marginally detected with the EVN at 1.7~GHz only, but~the strong side lobe patterns in the dirty images prevent us from locating the source with high confidence, and~brightness distribution models cannot be reliably fitted to the visibility data of these weak VLBI sources. This is typical for extended sources when only the shortest interferometer baselines detect the emission. We tentatively assume that the brightness peak of the dirty image coincides with the AGN~core.

\subsection{VLA~Data}

To determine the VLASS flux densities, we loaded the Quick Look images covering our target sources into AIPS. We used the task {\sc JMFIT} to fit a circular Gaussian to the image data at the given positions (see~Table~\ref{tbl:sources}). The~typical brightness uncertainty of the VLASS images is $\approx 15 \%$ \cite{vlassmemo}. We~adopted this as the error of the fitted 2.7-GHz flux density values that are given in Table~\ref{tbl:props}, along~with the fitted 1.7- and 5-GHz EVN flux densities in the case of J1026$-$0042. For~the remaining four LLAGNs considered to be marginally detected at 1.7~GHz and not detected at 5~GHz with the EVN, at~the latter frequency, we give flux density upper limits for any compact radio source of the restoring beam size of about 1.5 mas $\times$ 2.5 mas, based~on the 7 $\sigma$ image noise level. At~1.7~GHz, we can obtain a flux density estimate of the radio emission that is mostly resolved out by the EVN, by~means of fitting a circular Gaussian model at the brightness peak location in the extended-field dirty maps. These models should be treated with caution and considered as tentative only. However, as~we show below,    the~flux density values determined this way are consistent with the VLA data measured at close-by~frequencies.

\begin{table}[H]
\caption{Flux densities with their associated uncertainties and the derived  $\alpha^{2.7}_{1.4}$ spectral indices of the target~sources.} 
\label{tbl:props}
\centering
\begin{tabular}{ccccccc}
\toprule
\multirow{2}{*}{\textbf{ID}} & \boldmath{S$_{1.4\,\rm{GHz}}$} & \boldmath{$S_{1.7\,\rm{GHz}}$} & \boldmath{$S_{2.7\,\rm{GHz}}$} & \boldmath{$S_{5\,\rm{GHz}}$} & \boldmath{$S_{8.4\,\rm{GHz}}$} & \multirow{2}{*}{\boldmath{$\alpha^{2.7}_{1.4}$ }} \\
  & \textbf{mJy} & \textbf{mJy} & \textbf{mJy} & \textbf{mJy} & \textbf{mJy} &  \\ 
\midrule
 SDSS J0741$+$3020 & 1.12 $\pm$ 0.16 & 0.28 $\pm$ 0.03 & 0.79 $\pm$ 0.12 &	$<$0.08 & 2.97 $\pm$ 0.01  & $-$0.53 $\pm$ 0.04 \\
 SDSS J0956$+$5735 & 0.88 $\pm$ 0.15 & 0.71 $\pm$ 0.07 & 0.47 $\pm$ 0.07 &	$<$0.08 & 2.88 $\pm$ 0.01  & $-$0.95 $\pm$ 0.03 \\
 SDSS J1014$+$0244 & 2.02 $\pm$ 0.13 & 1.70 $\pm$ 0.17 & 1.20 $\pm$ 0.18 &	$<$0.11 & 4.53 $\pm$ 0.02  & $-$0.80 $\pm$ 0.07 \\
 SDSS J1026$-$0042 & 1.85 $\pm$	0.15 & 1.24 $\pm$ 0.12 & 1.63 $\pm$	0.24 & 0.20 $\pm$ 0.02 & 3.68 $\pm$ 0.02  & $-$0.20 $\pm$ 0.09 \\
 SDSS J1447$+$0211 & 0.76 $\pm$	0.15 & 0.23 $\pm$ 0.02 & 0.55 $\pm$ 0.08 &	$<$0.08 & 2.76 $\pm$ 0.01  & $-$0.49 $\pm$ 0.08 \\
\bottomrule
\end{tabular}\\
\begin{tabular}{@{}c@{}} 
\multicolumn{1}{p{\linewidth-1.5cm}}{\footnotesize Notes: Column  2,  VLA~\cite{becker1995,lalandho}; Column  3, EVN (this paper); Column  4,      VLA~\cite{lacy2020}; Column  5, EVN (this paper); Column  6, VLA~\cite{lalandho}; Column  7, two-point spectral index calculated from the arcsec-scale VLA (1.4-GHz FIRST and 2.7-GHz VLASS) measurements~\cite{becker1995,lacy2020}}
\end{tabular}
\end{table}

\section{Discussion}
\label{s:discussion}
\unskip

\subsection{VLBI Structure of the Target~Sources}
\label{ss:results-evn}

\textls[-25]{From the five targeted LLAGN sources, only SDSS~J1026$-$0042 was firmly detected with the EVN. One source component was found at both observing frequencies, 1.7 and 5~GHz. The~VLBI positions (the coordinates corresponding to the origin of the maps in Figure~\ref{fig:J1026}) agree with each other within the uncertainty of about 1~mas. At~both frequencies, the~component is seen as a compact source. We~calculated the minimum resolvable angular size with the interferometer using Equation~(2) of~\cite{kovalev2005}. The~single detected component in SDSS~J1026$-$0042 is resolved at both frequencies as the fitted circular Gaussian diameters, 0.79 $\pm$ 0.06~mas and 0.065 $\pm$ 0.007~mas (FWHM) at 1.7 and 5 GHz, respectively, are larger than the minimum resolvable sizes. The~size uncertainties were calculated following \citet{fomalont1999}. Using the AIPS task MAXFIT, the~peak position of the 5-GHz image is at right ascension $\alpha_{\rm EVN} = 10^{\rm h} 26^{\rm m} 40.4381^{\rm s}$ and declination $\delta_{\rm EVN} = -00^\circ 42^\prime 06.496^{\prime\prime}$. In~comparison, the~most accurate optical position taken from the {\em Gaia} Data Release 2~\cite{gaia2016, gaia2018} database ({\url{https://gea.esac.esa.int/archive/}}) is $\alpha_{\rm Gaia} = 10^{\rm h} 26^{\rm m} 40.4371^{\rm s} \pm 0.0003^{\rm s}$, $\delta_{\rm Gaia} =-00^\circ 42^\prime 06.4843^{\prime\prime} \pm 0.0081^{\prime\prime}$. There is a radio--optical positional difference of $\sim 15$  mas in right ascension and $\sim 10$~mas in declination (Figure      \ref{fig:J1026}). Given~that the EVN positions are accurate to within 1~mas, and~for the optical position the uncertainties are $\Delta \alpha \approx 5$~mas and $\Delta \delta \approx 8$~mas, the~difference is significant. At~$z = 0.364$, this corresponds to ($\sim$90 $\pm$ 40)~pc projected linear distance between the {\em Gaia} optical and the VLBI radio position. According to \citet{plavin2019}, at~least 20--50 pc projected difference between the VLBI and {\em Gaia} positions is quite common in bright AGNs. They also found that in $80\%$ of Seyfert 2 galaxies, the~associated {\em Gaia} position related to the jet is located farther away from the nucleus than the VLBI core. This offset could be explained by the presence of a bright extended optical jet which shifts the {\em Gaia} centroid~\cite{kovalev2017, petrovkovalev2017}.}
While this interpretation might seem appealing in view of the extended lobe structure reported by \citet{lalandho}, in~our VLBI data,  we do not find any supporting evidence for large scale-jets, which makes this interpretation less~likely.

\textls[-15]{Based on our EVN measurements, we derived the observed brightness temperatures of SDSS~J1026$-$0042 at both frequencies using Equation     \eqref{eq:tb}  \cite{condon1982}. Here, $z$ is the redshift, $S_\nu$ is the flux density (measured in Jy) at $\nu$ frequency (GHz), and~$\theta$ is the FWHM of the fitted circular Gaussian model component (mas).}
\begin{equation} \label{eq:tb}
    T_{\rm{b,obs}} = 1.22 \times 10^{12} (1 + z) \frac{S_\nu}{\theta^2 \nu^2} \quad [\rm{K}]
\end{equation}

The derived values are $T_{\rm{b,obs}} = (1.15 \pm 0.22) \times 10^{9}$~K at 1.7 GHz, and~$T_{\rm{b,obs}} = (3.01 \pm 0.74) \times 10^{9}$~K at 5 GHz. The~values $T_{\rm{b,obs}} > 10^6$~K indicate non-thermal emission originating from AGN (e.g., \cite{kewley2000}). We also derived the $K$-corrected monochromatic radio power values for the compact VLBI-detected components of SDSS~J1026$-$0042: $P_{\rm{1.7GHz}}$ = (7.08 $\pm$ 0.48) $\times$ 10$^{23}$~W\,Hz$^{-1}$ and $P_{\rm{5GHz}}$ = (1.10 $\pm$ 0.13) $\times$ 10$^{23}$~W\,Hz$^{-1}$.
For~this calculation, we used Equation     \eqref{eq:luminosity}, where $D_{\rm{L}}$ is the luminosity distance and $\alpha = -1.73$ is the spectral index between 1.7 and 5~GHz derived from our VLBI model component flux densities (Table~\ref{tbl:props}).
\begin{equation} \label{eq:luminosity}
    P_{\nu} = 4 \pi D_{\rm{L}}^2 S_\nu (1+z)^{-1-\alpha}  
\end{equation}

With these values at hand, we can assess whether the radio emission could originate from supernova remnants in the host galaxy~\cite{kewley2000}. The~supernova rate corresponding to our radio powers ($\nu_{\rm{SN}}$ = 149 and 52~yr$^{-1}$ at 1.7 and 5 GHz, respectively) exceeds the rates of the most powerful starburst galaxies (e.g.,~\cite{mannucci2003, bondi2012}). An~individual supernova remnant is also ruled out, and~complexes of supernova remnants do not concentrate at mas scales. A~recent study investigating radio-quiet AGNs at $z < 0.8$, with~4 $\times$ 10$^{21} <$ P$_{\rm{1.4GHz}} <$ 7 $\times$ 10$^{24}$~W Hz$^{-1}$, and~L$_{\rm{bol}}$ $<$ 10$^{38}$~W~\cite{zakamska2016} concluded that the radio emission originating from these sources is dominated by AGN activity and not the host galaxy. Similar results were found by \citet{herrera2016} and \citet{maini2016}.

\subsection{The Marginally Detected~Sources}
\label{ss:results-detection}

As mentioned in Section \ref{s:datared}, we marginally detected four target sources in the 1.7-GHz EVN dirty images. This suggests the presence of some extended radio emission that is, however, completely resolved out on the long baselines. On~the other hand, these sources remained undetected at 5~GHz. 

\textls[-25]{We tentatively locate them at the respective brightness peaks of the phase-referenced EVN dirty maps. These coordinates are given in Table~\ref{tbl:marginal}, along with a radio--optical angular separation based on the optical coordinates in SDSS DR15~\cite{strauss2002,sdssdr15} (see Table~\ref{tbl:sources}). Unfortunately, unlike for SDSS~J1026$-$0042, there are no accurate {\em Gaia} optical positions available for these LLAGNs. The~SDSS positions are known to be accurate to $\sim$60~mas (1 $\sigma$) in both right ascension and declination~\cite{oroszfrey2013},~but~\citet{skipper2018} found an excess of nearby ($z<0.2$) extragalactic sources with SDSS--radio positional separations larger than $\sim$150~mas, which~may be caused by various effects. Therefore,   our tentative radio positions seem to be consistent with the optical ones within the rather large uncertainties. More sensitive VLBI observations, possibly involving baselines providing intermediate ($\sim$100-mas) angular resolution would be needed for reliable detection and imaging of these four sources, and~for determining their accurate radio~positions. }

\begin{table}[H]
\caption{Tentative VLBI positions of the sources marginally detected with the EVN at 1.7~GHz, and~the radio--optical positional~offsets.}  
\label{tbl:marginal}
\centering 
\begin{tabular}{cccccccc}
\toprule
\multirow{2}{*}{\textbf{ID}} & \multicolumn{3}{c}{\textbf{Right Ascension}} & \multicolumn{3}{c}{\textbf{Declination}} & \boldmath{$\Delta_{\rm radio-opt}$}  \\
   & \textbf{h} & \textbf{min} & \textbf{s} & \boldmath{$^\circ$} & \boldmath{$^\prime$} & \boldmath{$^{\prime\prime}$}& \textbf{mas}  \\
\midrule
 SDSS~J0741$+$3020 & 07 & 41 & 30.518 & $+$30 & 20 & 05.245 & 98 \\
 SDSS~J0956$+$5735 & 09 & 56 & 29.046 & $+$57 & 35 & 08.751 & 186 \\
 SDSS~J1014$+$0244 & 10 & 14 & 03.513 & $+$02 & 44 & 16.265 & 310 \\
 SDSS~J1447$+$0211 & 14 & 47 & 11.283 & $+$02 & 11 & 36.389 & 203 \\
\bottomrule
\end{tabular}
\end{table}
\unskip

\subsection{Our Target Sources in the~VLASS}
\label{ss:results-vlass}

The VLASS data that became available recently~\cite{lacy2020} allowed us to estimate the radio spectral indices of the target sources, independently of \citet{lalandho}. Four of the five sources have been clearly detected with the VLASS at the first epoch (Table~\ref{tbl:props}). In~the case of SDSS~J0956$+$5735, there appears a $\sim$4 $\sigma$ peak coinciding with the optical position, suggesting the presence of a radio source. In~the future, in~the completed VLASS with three  epochs combined, this source should also become securely detected. Apart from the poorer sensitivity, the~angular resolution of the 2.7-GHz VLASS images is about  three   times lower than that of the 8.4-GHz VLA images presented by \citet{lalandho}, making the direct comparison of the arcsec-scale source structures difficult. Nonetheless, the~VLASS images indicate single components and not symmetric double structures, which they might not be able to~resolve.

\subsection{Revised Radio Spectral~Indices} 
\label{ss:results-spectralindex}

Based on the newly-obtained flux density data, we reexamined the radio continuum spectral properties of the target LLAGN sources. Figure~\ref{fig:flux} shows their radio spectra between 1.4 and 8.4~GHz. For~recalculating the spectral indices, we used the FIRST 1.4-GHz~\cite{becker1995} and the VLASS 2.7-GHz flux densities (Table~\ref{tbl:props}), both~based on VLA measurements. The~resulting $\alpha^{\rm{2.7}}_{\rm{1.4}}$ values are given in Table~\ref{tbl:props} and Figure~\ref{fig:flux}. We excluded from the spectral fit the 8.4-GHz VLA values reported by \citet{lalandho} since those appear in most cases an order of magnitude higher than expected from the trend based on lower-frequency data.  We did not include the VLBI flux density values either, because~the EVN provides about three orders of magnitude higher resolution than the VLA, thus probing radio emission on a much smaller spatial scale, detecting mas-scale compact emission only. Moreover, four out of   five sources are undetected at 5~GHz with the EVN, leading to flux density upper limit estimates~only.

\begin{figure}[H]
\centering
 \includegraphics[width=13cm]{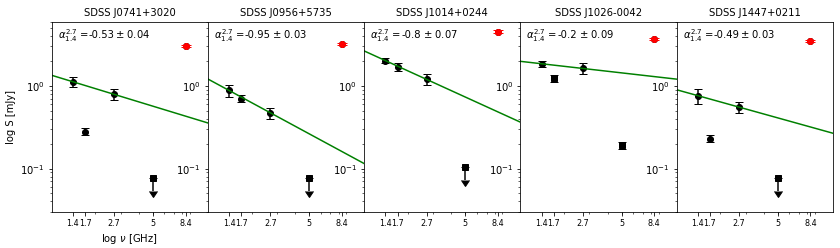}
 \caption{The continuum radio spectra of the sources between 1.4 and 5~GHz with flux density values and their associated uncertainties taken from Table~\ref{tbl:props} (black dots). For~sources undetected with the EVN, flux density values at 5~GHz are shown as upper limits indicated by arrows. For~SDSS~J1026$-$0042 detected with the EVN, among~the other 1.7 GHz VLBI flux densities, the~values should in fact be considered as lower limits since any extended emission is resolved out on the long baselines. The~green lines represent power-law spectra calculated from two VLA data points, FIRST at 1.4~GHz and VLASS at 2.7-GHz. The~corresponding $\alpha^{\rm{2.7}}_{\rm{1.4}}$ spectral index values are displayed in the top left corner of the panels. For~comparison, the~8.4-GHz VLA flux densities taken from \citet{lalandho} are also indicated with red symbols. In~all cases, these are much higher than expected from the extrapolation of 1.4- and 2.7-GHz VLA values, indicating an issue with the values derived in~\cite{lalandho}.} 
 \label{fig:flux}
\end{figure}

Considering our $\alpha^{\rm{2.7}}_{\rm{1.4}}$ spectral index values derived from VLA data (Table~\ref{tbl:props}), the~only source that has a flat radio spectrum is SDSS~J1026$-$0042 ($\alpha^{\rm{2.7}}_{\rm{1.4}} = -0.20$). All others can be classified as steep-spectrum sources with spectral index below $\approx -0.5$. Notably, this finding is perfectly consistent with the fact that the only flat-spectrum source, SDSS~J1026$-$0042, shows a compact VLBI component, while the steep-spectrum ones are all almost completely resolved out with the EVN. However, the~$\alpha^{\rm{8.4}}_{\rm{1.4}}$ spectral indices (Table~\ref{tbl:sources}) determined by \citet{lalandho} are markedly different, with~opposite sign, all~indicating inverted spectra ($\alpha^{\rm{8.4}}_{\rm{1.4}}$ > 0.38). Higher flux densities at 8.4~GHz, as~seen in Figure~\ref{fig:flux}, could~in principle be caused by extreme variability, because~the measurements at different frequencies were not made simultaneously. However, it is remarkable that all five sources show highly elevated flux densities at 8.4~GHz, rendering the variability explanation very unlikely, even though selection at higher frequencies  such as  8.4~GHz would be biased to flat-spectrum, more variable~sources.

Interestingly, among~the lowest-power radio sources in the whole Type 2 quasar sample where our sources were selected from~\cite{lalandho}, the~vast majority appear to have inverted spectra (see their Figure~6). This is even more puzzling in comparison with similar radio studies of Type 2 quasars~\cite{zakamska2004,martinez2006} where~these objects tend to have steep ($\alpha < -0.5$) spectrum. The~steep spectra indicate extended radio emission, explaining why we could not detect compact cores in four LLAGNs. 

The compact VLBI component found in SDSS~J1026$-$0042 appears to have a steep spectrum ($\alpha^{\rm{5}}_{\rm{1.7}} = -1.73$). While this would not be unprecedented since steep-spectrum and even ultra-steep-spectrum ($\alpha < -1$) radio cores are often found in LLAGNs (e.g., \cite{argo2013, fu2015, kharb2017, kharb2020}), we should treat this result with caution. The~reason for the ill-determined core spectrum could be that the source is very weak, and~model fitting to the EVN visibility data could be affected by phase-reference losses (e.g., \cite{marti2010}), especially at the higher frequency. This may have led to the underestimation of the flux density at 5~GHz, and~consequently the core spectral~index.   


\section{Conclusions}
\label{s:conclusions}

In this paper, we present   1.7- and 5-GHz EVN observations of five LLAGNs selected by their spectral (inverted-spectrum) and morphological (double-lobed symmetric) properties from~\cite{lalandho}. Out of the five sources, only one (SDSS~J1026$-$0042) was firmly detected with the EVN at both frequencies. It has a compact mas-scale VLBI core. 
The compact radio emission is a result of AGN activity. The~other four sources (J0741$+$3020, J0956$+$5735, J1014$+$0244, and~J1447$+$0211) are marginally detected at 1.7~GHz but not detected at 5~GHz with the EVN. It is consistent with dominantly extended, steep-spectrum radio emission in these~objects. 

With the help of VLASS data, we revised the radio spectral index for our five target sources. We~found that they have flat or steep spectra between 1.4 and 2.7~GHz, contrary to the inverted spectra reported earlier~\cite{lalandho}. The~revised overall spectral properties are consistent with the outcome of the EVN observations, as~the only flat-spectrum source (SDSS~J1026$-$0042) contains a compact VLBI core. Based~on our results, we suspect that spectral indices for at least some other inverted-spectrum  LLAGNs found by \citet{lalandho} may need to be revised. This could be done using new survey data becoming available  such as those from the~VLASS.

Because of the dubious sample selection, the~applicability of our proposed method to find kpc-scale dual AGNs by targeting truly inverted-spectrum arcsec-scale symmetric double extragalactic radio sources with high-resolution VLBI imaging observations remains to be~proven.


\vspace{10pt}

\authorcontributions{Conceptualization  and  investigation, Z.P., R.D.,  and~S.F.;  Formal analysis, M.K. and  S.F.; Writing---Original draft preparation, M.K. and  S.F.;  and writing--review and editing, Z.P. and~R.D. All authors have read and agreed to the published version of the manuscript.}

\funding{The research leading to these results has received funding from the European Commission Seventh Framework Programme (FP/2007--2013) under grant agreement No. 283393 (RadioNet3). This work is part of the project ``Transient Astrophysical Objects'' GINOP 2.3.2-15-2016-00033 of the National Research, Development and Innovation Office (NKFIH), Hungary, funded by the European Union.}

\acknowledgments{\textls[-15]{The EVN is a joint facility of independent European, African, Asian, and~North American radio astronomy institutes. Scientific results from data presented in this publication were derived from the project EP093. The~National Radio Astronomy Observatory is a facility of the National Science Foundation operated under cooperative agreement by Associated Universities, Inc. This research has made use of the NASA/IPAC Extragalactic Database (NED), which is funded by the National Aeronautics and Space Administration and operated by the California Institute of~Technology.
Funding for the SDSS and SDSS-II has been provided by the Alfred P. Sloan Foundation, the~Participating Institutions, the~National Science Foundation, the~U.S. Department of Energy, the~National Aeronautics and Space Administration, the~Japanese Monbukagakusho, the~Max Planck Society, and~the Higher Education Funding Council for England. The~SDSS Web Site is \url{http://www.sdss.org/}. The~SDSS is managed by the Astrophysical Research Consortium for the Participating Institutions. The~Participating Institutions are the American Museum of Natural History, Astrophysical Institute Potsdam, University of Basel, University of Cambridge, Case Western Reserve University, University of Chicago, Drexel University, Fermilab, the~Institute for Advanced Study, the~Japan Participation Group, Johns Hopkins University, the~Joint Institute for Nuclear Astrophysics, the~Kavli Institute for Particle Astrophysics and Cosmology, the~Korean Scientist Group, the~Chinese Academy of Sciences (LAMOST), Los Alamos National Laboratory, the~Max-Planck-Institute for Astronomy (MPIA), the~Max-Planck-Institute for Astrophysics (MPA), New Mexico State University, Ohio State University, University of Pittsburgh, University of Portsmouth, Princeton University, the~United States Naval Observatory, and~the University of~Washington. 
This work presents results from the European Space Agency (ESA) space mission {\em Gaia}. {\em Gaia} data are being processed by the {\em Gaia} Data Processing and Analysis Consortium (DPAC). Funding for the DPAC is provided by national institutions, in~particular the institutions participating in the {\em Gaia} MultiLateral Agreement (MLA). The~{\em Gaia} mission website is \url{https://www.cosmos.esa.int/gaia}. The~{\em Gaia} archive website is \url{https://archives.esac.esa.int/gaia.}}}

\conflictsofinterest{The authors declare no conflict of~interest.} 

\abbreviations{The following abbreviations are used in this manuscript:\\}

\noindent 
\begin{tabular}{@{}ll}
AGN & active galactic nucleus\\ 
AIPS & NRAO Astronomical Image Processing System\\
DR & data release\\
EVN & European VLBI Network\\
FIRST & Faint Images of the Radio Sky at Twenty Centimeters\\
FWHM & full width at half-maximum\\
HPBW & half-power beam width\\
JIVE & Joint Institute for VLBI European Research Infrastructure Consortium\\
LINER &  low-ionization nuclear emission-line region\\
LLAGN & low-luminosity AGN\\
NRAO & U.S. National Radio Astronomy Observatory \\
NVSS & NRAO VLA Sky Survey \\
SDSS & Sloan Digital Sky Survey\\
SKA & Square Kilometre Array\\
SNR & signal-to-noise ratio \\
ULIRG & ultraluminous infrared galaxy \\
VLA & Karl G. Jansky Very Large Array\\
VLASS & Karl G. Jansky Very Large Array Sky Survey\\
VLBI & very long baseline interferometry
\end{tabular}}

\reftitle{References}




\end{document}